\documentclass[11pt]{article}

\usepackage{mathptmx}
\usepackage{caption}

\usepackage{amsmath,amssymb,amsfonts,amsthm} 
\usepackage{mathrsfs,latexsym} 

\DeclareMathAlphabet{\pazocal}{OMS}{zplm}{m}{n}

\usepackage{color}
\usepackage{cite}
\usepackage{epsfig}
\usepackage{graphicx}

\usepackage{bm}

\usepackage{cite}\medskip

\numberwithin{equation}{section}

\newcommand{\CC}{{\mathbb C}}
\newcommand{\RR}{{\mathbb R}}

\newcommand{\NN}{{\mathbb N}}
\newcommand{\TT}{{\mathbb T}}

\newcommand{\Dc}{{\mathcal{D}}}
\newcommand{\Ec}{{\mathcal{E}}}
\newcommand{\Fc}{{\mathcal{F}}}

\newcommand{\Lc}{{\mathcal{L}}}
\newcommand{\Mc}{{\mathcal{M}}}

\newcommand{\Oc}{{\mathcal{O}}}

\newcommand{\Pc}{{\mathcal{P}}}

\newcommand{\bix}{{\bm x}}

\newcommand{\Alg}{\pazocal{A}}                   
\newcommand{\Bog}{\pazocal{B}}                   
\newcommand{\Test}{\pazocal{D}}                  
\newcommand{\Group}{\pazocal{G}}

\newcommand{\scirc}{\mbox{\footnotesize $\circ$}}
\newcommand{\supp}{{\mbox{supp} \, }}

\newcommand{\scale}[2]{\scalebox{#1}{#2}}       

\newcommand{\oproducts}[5]{\scale{#1}{$\bigcirc  
\hspace{\dimexpr #2pt \relax} \raisebox{\dimexpr #3pt \relax} 
{\scalebox{#4}{#5}}$}{}}           

\newcommand{\tpunkt}{\oproducts{1}{-6.9}{1.5}{0.5}{$\bullet$} \hspace{1pt}}
\newcommand{\mpunkt}{\raisebox{1.5pt}{\oproducts{0.7}{-6.9}{1.5}{0.5}{$\bullet$}}}

\newcommand{\tstern}{\oproducts{1}{-9}{-0.5}{1.3}{$\star$}  \hspace{1pt}}
\newcommand{\mstern}{\raisebox{1.5pt}{\oproducts{0.7}{-9}{0}{1.3}{$\star$}}}

\newcommand{\tzeit}{\oproducts{1}{-8.5}{0}{0.8}{$T$}  \hspace{1pt}}
\newcommand{\mzeit}{\raisebox{1.5pt}{\oproducts{0.7}{-8}{0}{0.75}{$T$}}}

\newcommand{\tnormal}{\oproducts{1}{-9.25}{0}{1}{$w$} \ }
\newcommand{\mnormal}{\raisebox{1.5pt}{\oproducts{0.8}{-9}{0}{1}{$w$}}}

\newcommand{\no}[1]{:\! \! #1 \! \!:}

\newcommand{\ad}[1]{\mbox{Ad} \, #1 }

\def\ie{{\it i.e.\ }}
\def\viz{{\it viz.\ }}
\def\etc{{\it etc}}

\def\change{\marginpar[\hfill !?]{!?}}

\def\primal{principal\ }

\begin{document} 

%

\title{A C*-algebraic approach \\ to interacting quantum field theories \\
{\large In memory of Eyvind H.\ Wichmann} }
\author{\large Detlev Buchholz${}^{(1)}$ \ and 
\ Klaus Fredenhagen${}^{(2)}$ \\[3mm]
\small 
${}^{(1)}$ Mathematisches Institut, Universit\"at G\"ottingen, \\
\small Bunsenstr.\ 3-5, 37073 G\"ottingen, Germany\\[5pt]
\small
${}^{(2)}$ 
II. Institut f\"ur Theoretische Physik, Universit\"at Hamburg \\
\small Luruper Chaussee 149, 22761 Hamburg, Germany \\
}
\date{}

\maketitle

{\small 
  \noindent {\bf Abstract.}
A novel C*-algebraic framework is presented for relativistic
quantum field theories, fixed by a 
Lagrangean. It combines the postulates of local quantum physics,
encoded in the Haag-Kastler axioms, with insights gained in the perturbative
approach to quantum field theory. Key ingredients are an appropriate version
of Bogolubov's relative $S$-operators and a reformulation of the
Schwinger-Dyson equations. These are used to define for any classical
relativistic Lagrangean of a scalar field 
a non-trivial local net of C*-algebras, encoding
the resulting interactions at the quantum level. The construction
works in any number of space-time dimensions. It reduces the longstanding
existence problem of interacting quantum field theories in physical spacetime
to the question of whether the C*-algebras so constructed admit 
suitable states, such as stable ground and equilibrium states. 
The method is illustrated on the example of a non-interacting field
and it is shown how to pass from it 
within the algebra to interacting theories by relying on
a rigorous local version of the interaction picture.  } \\[2mm]
 \\[1mm]

\section{Introduction}
\setcounter{equation}{0}

Quantum field theory aims to reconcile the principles of quantum physics,
governing the microcosmos, with those of relativistic causality, regulating 
all physical processes. It was conceived immediately after the advent of
quantum mechanics as a framework for the quantization
of the electromagnetic field. 
Yet, whereas quantum mechanics quickly matured into a meaningful theory with
solid mathematical foundations, the consolidation of quantum field theory
took several decades and, as a matter of fact, has not yet come to a fully
satisfactory end. In the course of these endeavors it became clear that
the framework of quantum field theory reaches far beyond electromagnetism.
In fact, it covers all fundamental forces known to date which,
with the exception of gravity, are subsumed in the standard model
of particle physics. 

\medskip 
On the mathematical side there exist two complementary attempts 
towards mastering the theory. With regard to its computational aspects,
one proceeds usually from a specific classical Lagrangean, encoding
the postulated field content and its interactions. One then
``quantizes'' the
theory, commonly by writing down path integrals or relying on 
canonical quantization schemes. This 
informal starting point acquires some precise meaning in the form of
calculational rules, ranging from Feynman graphs in renormalized 
perturbation theory to lattice approximations. It 
leads to a multitude of theoretical predictions
which are in solid agreement with experimental results. 
Yet the mathematical status of the starting point, \ie the existence 
of the conceived quantized theory, is not touched upon by these 
investigations and, most likely, cannot be clarified in
this manner. 

\medskip 
It is the latter issue which is in the focus of  
attempts to put quantum field theory on firm mathematical 
grounds. There one proceeds from the conceptual foundations 
of the theory, such as its probabilistic interpretation 
and its causal structure, and casts them into 
proper mathematical conditions. In this manner one obtains a      
general framework for quantum field theory, such as the Wightman 
axioms and their Euclidean ramifications or the Haag-Kastler 
postulates of algebraic quantum field theory \cite{Haag92,Haag64}. 
These settings 
have been the basis for the explanation of distinctive 
features of particle physics, such as the possible manifestations 
of particle statistics, the existence of anti-particles and the 
appearance of internal symmetry groups. Moreover, they form the arena for 
the rigorous construction of quantum field theoretic models. 
Yet, disregarding examples in a low dimensional model world or 
non-interacting theories, these constructive attempts  
have not yet succeeded in establishing the existence 
of quantum field theories in real spacetime, which comply with 
all basic constraints put forward in the general framework  
\cite{GJ85,Summers12}.  

\medskip 
It is the aim of the present article to combine these
two attempts. Our construction relies on insights gained in a 
perturbative approach to quantum field theory, 
which can be traced back to some seminal work of Bogolubov
\cite{BP57,BS59}.
The essential ingredient in this approach are unitary $S$-operators, 
which may be regarded as local versions of a scattering matrix. In 
order to pass from one theory to another one proceeds from them to 
relative $S$-operators, depending on the interaction. The latter unitary 
operators satisfy causality relations which are model independent. 
Moreover, for given Lagrangean, they allow to describe the effect of 
local changes of the underlying field by corresponding 
variations of the action, fixed by the 
Lagrangean. This feature is closely related to  
the Schwinger-Dyson equations, which comprise the equations of 
motion of the theory. In the simple closed form given here, 
the presentation of these equations seems to be new. 

\medskip 
The relative $S$-operators are constructed in the perturbative 
approach as formal power 
series, leaving aside questions of convergence. In lieu thereof we 
introduce in the present article 
a unitary group that is generated by abstract $S$-operators,  
encoding the above-mentioned causal and dynamical constraints. 
These unitaries are labelled by local functionals, mapping classical field
configurations into real numbers. In order to simplify the discussion, 
we restrict our attention to $d$-dimensional Minkowski space which 
carries a scalar field, being described by smooth, real-valued  
functions. The functionals which we consider are determined by 
polynomials formed out of the field and its derivatives, which are   
integrated with test functions. Let us emphasize that we are not 
introducing ``quantization rules'' for the underlying classical 
theory. The classical theory primarily serves 
to describe the localization properties of the $S$-operators
and to indicate which particular observable   
we have in mind, without trying to specify its concrete quantum 
realization. Thus, in accord with the doctrine of Niels Bohr, 
we are using ``common language'' in order to describe  
observables and operations relating to the quantum world. 

\medskip 
Making use of a standard construction method, we shall extend the 
unitary group so defined to a C*-algebra. This algebra
is shown to be the inductive limit of a local net of algebras on 
Minkowski space which comply with the condition of locality
(Einstein causality). Moreover, the spacetime symmetry 
group, the Poincar\'e group, acts by automorphisms on this
net, in accordance with the Haag-Kastler postulates. 
Having established the general framework, we will illustrate its 
usefulness by considering the algebra determined by the Lagrangean 
of a non-interacting field. It turns out that it contains the
Weyl operators of a free field, satisfying the Klein-Gordon equation and
having c-number commutation relations. This proves that the algebra
is non-trivial and 
encodes specific dynamical information. We will therefore refer to 
it as ``dynamical algebra''. We then discuss 
the case of interacting theories and show that the corresponding
operators are related to those of the non-interacting theory by the adjoint 
action of $S$-operators which involve functionals describing 
the suitably localized interaction. This result 
justifies within the present setting the interpretation of the $S$-operators 
as localized scattering matrices. 

\medskip
The dynamical algebra has all properties which are needed to identify 
vacuum states or thermal equilibrium states in its dual space. These are 
commonly taken as characteristics for the stability of the theory. The 
question of whether 
such states exist is expected to depend on the form of the 
Lagrangean entering in the definition of the underlying group and the 
dimension $d$ of Minkowski space. As a matter of fact, the existence of such 
states may not always be expected in theories of physical interest, 
such as in massless theories in low spacetime dimensions;
there one has to rely on milder stability conditions. The existence 
of vacuum and thermal equilibrium states in physical 
spacetime has been established in interacting theories in the 
perturbative approach to the $S$-operators \cite{EG,FL14,Duch}. But 
these encouraging results 
do not yet settle the problem for the dynamical algebras in the present 
C*-algebraic setting. An affirmative solution would be a vital step in 
the consolidation of the mathematical foundations of quantum field theory. 

\medskip
Our article is organized as follows. In the subsequent section we 
introduce the concepts used in the framework of classical field 
theory, which enter in our construction. Section 3 contains the 
definition of the dynamical algebra and the discussion of its general 
properties. In Sect.\ 4 we elaborate on the case of non-interacting 
theories and in Sect.\ 5 on theories involving interactions. The
article concludes with a summary and outlook on generalizations of the 
present results. In the appendix, the dynamical 
relations used in our approach are derived from the Schwinger-Dyson 
equations. 

\section{Classical field theory}
\setcounter{equation}{0}

In order to simplify the discussion, we restrict our attention
to fields on Minkowski space; yet the present framwork can be 
extended to fields on arbitrary Lorentzian manifolds. 
So let $\Mc$ be $d$-dimensional Minkowski space with its standard 
metric $g(x,x) \doteq x_0^2 - \bix^2$,
where $x_0, \bix$ denote the time and space components of 
$x \in \RR^d$. The symmetry group of $\Mc$ is the Poincar\'e group
\mbox{$\Pc = \RR^d \rtimes \Lc$}, consisting of the semi-direct product of 
translations and (proper, orthochronous) Lorentz transformations. 

\medskip 
We consider a scalar field on $\Mc$. Its configuration space 
$\Ec$ is the real vector 
space of smooth functions $\phi: \Mc \rightarrow \RR$ on which 
the Poincar\'e transformations $P \in \Pc$ act by 
automorphisms, $P \phi(\, \cdot \,) \doteq \phi(P \, \cdot \,)$.
Note that the field is not assumed to satisfy some field equation,
it is ``off shell" according to standard terminology.

\medskip 
For the sake of simplicity, we restrict our 
attention to functionals 
\mbox{$F : \Ec \rightarrow \RR$} \ of the specific form 
$$
F[\phi] = \int \! dx \, 
\sum_{n = 0}^{N} \, g_n(x) \, \phi(x)^n  \, ,
$$
where $g_n \in \Test(\RR^d)$ are arbitrary 
test functions. If $N > 2$, the sum 
contains terms describing some self-interaction of the field.
The resulting space $\Fc$ is sufficiently big in order to deal with 
the Lagrangeans of interest here, cf.~below. Moreover, given any field
$\phi_0 \in \Ec$, $\Fc$ is stable under the shifts   
$F \mapsto F^{\, \phi_0}$, defined by 
$F^{\, \phi_0}[\phi] \doteq F[\phi + \phi_0]$, $\phi \in \Ec$.
Wheras the functionals $F$ are in general not linear, one easily checks that 
they satisfy the additivity relation 
$$
F[\phi_1 + \phi_2 + \phi_3] = F[\phi_1 + \phi_3] - F[\phi_3] +
F[\phi_2 + \phi_3]
$$
for arbitrary $\phi_3$, provided the supports of the fields 
$\phi_1$ and $\phi_2$ are disjoint. This feature is a consequence of the 
locality properties of the functionals. 

\medskip
The support of functionals on $\Ec$ can be intrinsically defined \cite{BDF09}.
For the present family of functionals $F \in \Fc$, 
it can be identified with the 
union of the supports of the underlying test functions $g_n$
for $n \geq 1$.
The action of the Poincar\'e transformations 
\mbox{$P \in \Pc$} on $\Ec$ can be
transferred to the functionals $\Fc$ by shifting them to the 
underlying test functions,
$$
F_P[\phi] \doteq F[P \phi] = \int \! dx \, 
\sum_{n = 0}^{N} \, g_n(P^{-1}x) \, \phi(x)^n  \, . 
$$
Thus, if $F$ has support in some region $\Oc \subset \Mc$,
then $F_P$ has support in $P \, \Oc$.

\medskip
The Lagrangean densities 
of the field which we consider here have the customary form 
\begin{equation} \label{e2.1}
x \mapsto L(x)[\phi] =  1/2 \, (\partial_\mu \phi(x) \, \partial^\mu \phi(x) 
- m^2 \, \phi(x)^2 ) - \sum_{n = 0}^{N} \, g_n(x) \, \phi(x)^n \, ,
\end{equation}
where $m \geq 0$ is the mass and 
$g_n \in \RR$ are fixed coupling constants. If $N > 2$,
they describe some self-interaction of the field. Other local 
interaction potentials can be treated in a similar manner. 
We regard these densities as  
distributions $L$ on the space of test functions
$\Test(\Mc)$ with values in functionals, \viz
$$
L(f)[\phi] \doteq \int \! dx \, f(x) \, L(x)[\phi] \, \in \, \RR \, , \quad 
f \in \Test(\Mc) \, , \ \phi \in \Ec \, .
$$ 
Given a Lagrangean density $L$, these integrals 
define localized versions of a corresponding
action, which informally corresponds to the constant function $f = 1$. 
In spite of the fact that we do not have at our disposal the full action,
field equations can be derived in the present setting in the sense 
of distributions by proceeding to relative actions. 
Denoting the subspace of compactly supported fields 
by $\Ec_0 \subset \Ec$, the family of relative actions 
fixed by $L$ consists of the maps 
$\delta L : \Ec_0 \times \Ec \rightarrow \RR$ given by 
$$
\delta L(\phi_0)[\phi] \doteq L(f_0)^{\phi_0}[\phi] - L(f_0)[\phi]
= L(f_0)[\phi + \phi_0] - L(f_0)[\phi] \, , \quad \phi_0 \in \Ec_0 \, ;
$$
the test function $f_0$ has to be equal to $1$ on the support
of $\phi_0$. Because of the local structure of the Lagrangean density,
the relative actions do not depend on the particular choice of 
$f_0$ satisfying this condition. Morover, they belong
to the space of functionals $\Fc$, defined above. This is so since 
the terms involving the kinetic energy of the field 
cancel each other 
and terms which are linear in derivatives of the field can 
be transformed into terms which are linear in the field by partial 
integration. So the relative actions depend only on powers of the 
field $\phi$.

\medskip
The derivative of a relative 
action with regard to $\phi_0$ defines the Euler-Lagrange 
derivative \, $\epsilon L : \Ec_0 \times \Ec \rightarrow \RR$, 
$$
 \epsilon L(\phi_0)[\phi]  
\doteq \frac{d}{du} \, \delta L(u \, \phi_0)[\phi] \Big|_{u = 0} \, .
$$
A field $\phi$ is said to satisfy the Euler-Lagrange equation in the sense 
of distributions, 
\ie it is ``on shell'', if $\epsilon L(\phi_0)[\phi] = 0$ for 
all $\phi_0 \in \Ec_0$.  

\section{The dynamical C*-algebra}
\setcounter{equation}{0}

We turn now to the construction of the dynamical C*-algebra and the 
discussion of its general properties. As already mentioned, 
this algebra has its conceptual roots in the perturbative approach 
to quantum field theory. We present here the essential elements of our 
approach; the underlying arguments, motivating the proposed structures, 
are explained in the appendix.

\medskip 
Given a Lagrangean $L$,
we construct in a first step a corresponding 
group $\Group_L$. Its elements are abstract $S$-operators 
$S(F)$, which are labelled by functionals $F \in \Fc$. 
As already mentioned, the functionals can be 
shifted by the fields $\phi_0 \in \Ec_0$, putting
$F^{\phi_0}[\phi] \doteq F[\phi + \phi_0]$,
$\phi \in \Ec$. Utilizing the localization properties
of the functionals,
we say that the support of a functional $F_1$ is later than that 
of $F_2$ if there exists some Cauchy surface $C$ such that the support 
of $F_1$ lies above and that of $F_2$ below that surface relative to 
the time orientation of $\Mc$. 
We also recall that 
$\delta L(\phi_0)$ denotes the relative action for given 
field $\phi_0 \in \Ec_0$.
With these ingredients, the group $\Group_L$ is defined as follows. 

\medskip 
\noindent \textbf{Definition:} \ Given a Lagrangean $L$, the corresponding
group $\Group_L$ is the free group generated by 
elements $S(F)$, $F \in \Fc$, modulo the relations \\[1mm]
(i) \ $S(F) \, S(\delta L(\phi_0)) = S(F^{\phi_0} + \delta L(\phi_0))
=  S(\delta L(\phi_0)) \, S(F)$ \ for \ $\phi_0 \in \Ec_0$, 
$F \in \Fc$, \\[1mm]
(ii) $\, S(F_1 + F_2 + F_3) = S(F_1 + F_3) \, S(F_3)^{-1} \, S(F_2 + F_3)$ 
for any $F_3 \in \Fc$, provided the support of $F_1 \in \Fc$ is later than that 
of $F_2 \in \Fc$. 

\begin{figure}[h]
\centering
\includegraphics[width=0.6\textwidth]{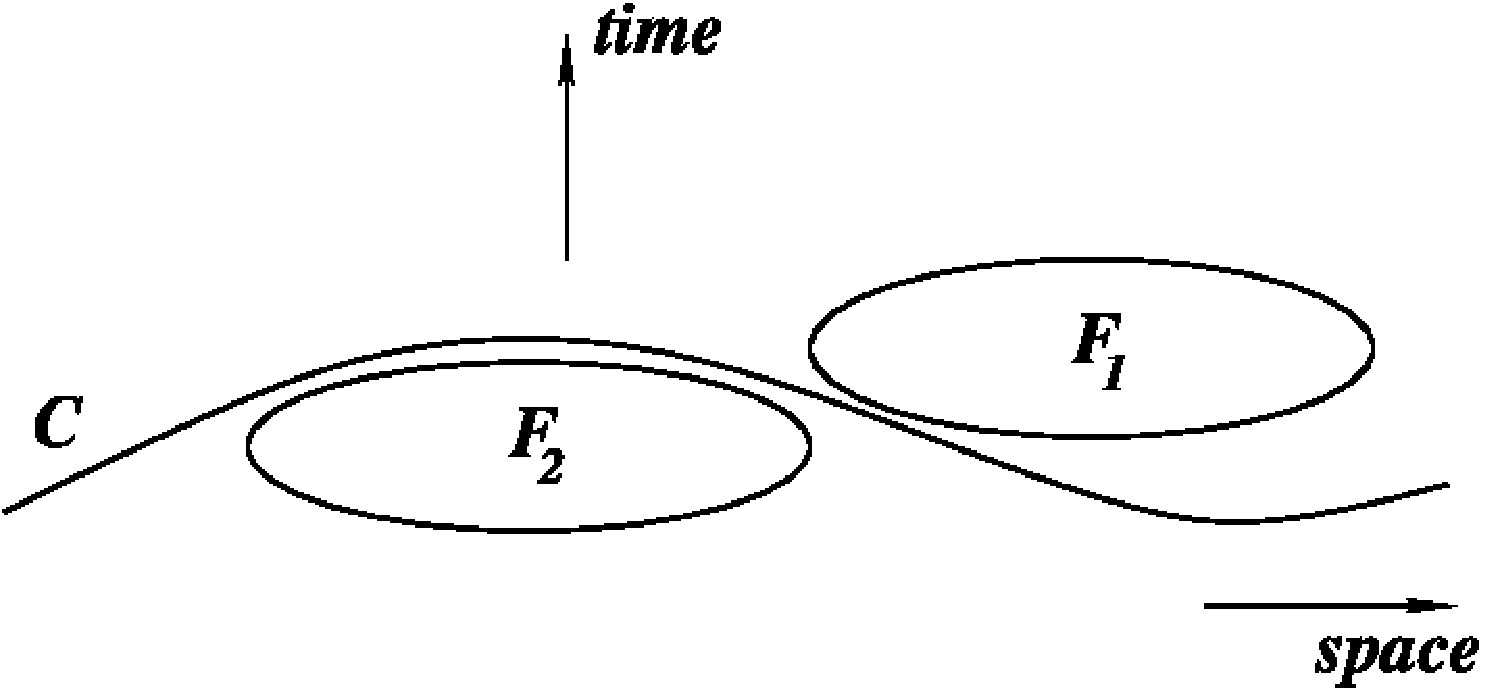}
\captionsetup{labelformat=empty}
\caption{\small 
Fig. \ The support  of functional $F_1$ is later than that of $F_2$}
\end{figure}

Relation (i) describes the dynamics incorporated  
in $\Group_L$. Putting $\phi_0 = 0$, one finds that $S(0) = 1$. 
The factorization relation (ii) comprises the causal 
properties of $\Group_L$. Putting
$S_3 = 0$, one obtains in particular 
$ S(F_1) \, S(F_2) = S(F_1 + F_2) $
if the support of $F_1$ is later than that of $F_2$. If the supports
of  $F_1$ and $F_2$ are spacelike separated, this condition implies that the 
corresponding elements commute since then there exist Cauchy surfaces 
separating the supports of the functionals in either temporal order,
cf.\ the figure below.  

\bigskip 
\begin{figure}[h]
\centering
\includegraphics[width=0.6\textwidth]{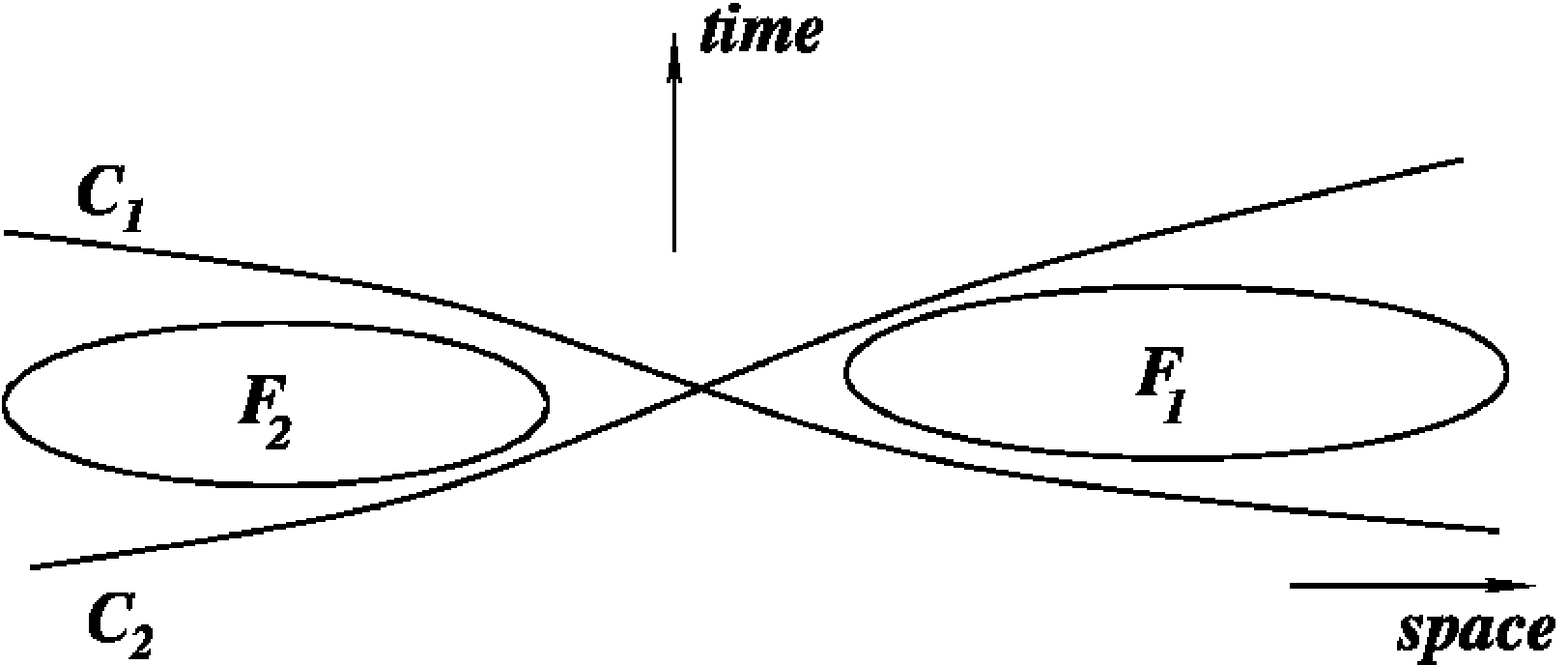}
\captionsetup{labelformat=empty}
\caption{Fig. \small \ Functionals $F_1$ and $F_2$ with 
spacelike separated supports}
\end{figure}

\medskip 
The preceding relations imply that the 
group $\Group_L$ has a center. A basic 
central subgroup is determined by the constant 
functionals $F_c$ which, for $c \in \RR$, are given by
$F_c[\phi] = c$, $\phi \in \Ec$. Since their  
support is empty, 
it follows from the causal factorization 
property of the $S$-operators 
that $S(F + F_c)= S(F)S(F_c) = S(F_c)S(F)$, $F \in \Fc$. 
Hence $c \mapsto S(F_c)$ defines a unitary 
representation of $\RR$ in the center of $\Group_L$.

\medskip
Another interesting subgroup in the center of $\Group_L$ is 
related to the dynamics. It is determined by the 
$S$-operators fixed by the relative actions, 
$S(\delta L(\phi_0))$, $\phi_0 \in \Ec_0$. They lie in the 
center according to the dynamical equations. To see that they 
form a group, note that according to these equations one has 
$$
S(\delta L(\phi_1)) \, S(\delta L(\phi_2)) =
S(\delta L(\phi_1)^{\phi_2} + \delta L(\phi_2))
= S(\delta L(\phi_1 + \phi_2)) \, , \quad 
\phi_1, \phi_2 \in \Ec_0 \, .
$$
These $S$-operators allow to discriminate off-shell from on-shell
fields. 

\medskip
We also note that 
$\Group_L$ is stable under the action of the Poincar\'e transformations
$P \in \Pc$, inducing the maps $S(F) \mapsto S(F_P)$. In case of
relation (ii), this is obvious since the causal order of 
the supports of functionals remains unaffected by the action of 
the elements of $\Pc$, which do not change the
time direction. With regard to relation~(i), this is a 
consequence of the equality 
$F^{\phi_0}{}_P[\phi] = F[P \phi + \phi_0]
= F_P^{\, P^{-1} \phi_0}[\phi]$ and  
the fact that the Lagrangean density transforms as a 
scalar under Lorentz transformations. It implies that  
$\delta L(\phi_0)_P[\phi] = \delta L(\phi_0)[P \phi] = 
\delta L(P^{-1}\phi_0)[\phi]$, \, 
$\phi \in \Ec$. 

\medskip
As is common practice, we use units, where Planck's constant 
has the value $1$. Since it is of interest to study the effect of 
hypothetical changes of this fundamental constant, we also
consider the groups, where this constant is scaled by some 
factor $h > 0$. It amounts to proceeding to the scaled 
Lagrangean $L_h \doteq h^{-1} L$ and scaled $S$-operators 
given by $S_h(F) \doteq S(h \, F)$, $F \in \Fc$.
This scaling neither affects the localization properties 
nor the Poincar\'e covariance of the functionals. 

\medskip 
We proceed now from $\Group_L$ to the corresponding group algebra 
$\Alg_L$ over $\CC$. This is a known procedure, which 
we briefly recall here. The algebra $\Alg_L$ 
is by definition the complex linear span of the elements $S \in \Group_L$. 
We also fix the central group elements $S(F_c)$, corresponding to the
constant functionals, putting  $S(F_c) = e^{ic} \, 1$, $c\in \RR$. 
The adjoint operators are defined by 
$(\sum c \, S)^* \doteq \sum \overline{c} \, S^{-1}$ and 
the multiplication in $\Alg_L$ is inherited from $\Group_L$ by the 
distributive law. 

\medskip 
On $\Alg_L$ there exists a functional $\omega$,
which is obtained by linear extension from the defining
equalities $\omega(S) = 0$ for $S \in \Group_L \backslash \{1\}$ and
$\omega(1) = 1$, cf.~\cite{BuCiRuVa}. 
Thus, for any choice of a finite number 
of different elements 
$S_i \in \Group_L$, $i = 1, \dots n$, one has 
$$
\omega\Big(\big(\sum_{i=1}^n  c_i S_i\big)^* 
\big(\sum_{j=1}^n  c_j S_j\big) \Big) 
= \sum_{i,j = 1}^n \overline{c}_i c_j \, \omega(S_i^{-1} S_j) = 
\sum_{i=1}^n \, |c_i|^2 \geq 0 \, .
$$
So, disregarding the zero element, the functional 
$\omega$ has positive values on positive operators in $\Alg_L$, 
\ie it is a faithful state. Whence, proceeding to the corresponding
GNS-representation, the operator norm of the elements of 
$\Alg_L$ in that representation defines a C*-norm on $\Alg_L$. We denote
by $\| \, \cdot \, \|$ the supremum of all C*-norms, obtained
in this manner by states on $\Alg_L$. (Note that this supremum
exists since each element of $\Alg_L$ is a finite sum of 
unitary operators.) Completing $\Alg_L$ in this norm topology,
we obtain a C*-algebra, which we denote by the same symbol.

\medskip \noindent
\textbf{Definition:} Given a Lagrangean $L$, the dynamical algebra $\Alg_L$
is the C*-algebra determined by the group $\Group_L$, as outlined above. 
 
\subsection{Haag-Kastler postulates} 

\noindent 
The dynamical algebra  $\Alg_L$ complies with the Haag-Kastler postulates 
of local quantum field theory \cite{Haag64} 
for any choice of Lagrangean $L$. In order
to verify this assertion, we first 
need to specify local subalgebras $\Alg_L(\Oc)$ for each bounded,  
causally closed spacetime region $\Oc \subset \Mc$. This is accomplished by 
making use of the support properties of the 
underlying functionals. We denote by $\Fc(\Oc) \subset \Fc$ the subspace
of all functionals having support in $\Oc$. It determines a corresponing 
subgroup $\Group_L(\Oc) \subset \Group_L$, generated by 
all $S(F)$ with $F \in \Fc(\Oc)$. From there one proceeds to the 
norm-closed subalgebra 
$$
\Alg_L(\Oc) \doteq \Big\{ \sum c \, S : \, c \in \CC \, , 
\  S \in \Group_L(\Oc) \Big\}^{\| \, \cdot \, \|} \subset 
\Alg_L \, , \quad \Oc \subset \Mc \, .
$$

By construction,  $\Alg_L(\Oc_1) \subset \Alg_L(\Oc_2)$
if $\Oc_1 \subset \Oc_2$, \ie the assignment $\Oc \mapsto \Alg_L(\Oc)$
satisfies the condition of isotony and thus defines a net of C*-algebras 
on~$\Mc$. Since all functionals $F \in \Fc$ underlying the construction of
$\Alg_L$ have compact supports, the C*-inductive
limit of this net coincides with $\Alg_L$. 

\medskip
Next, if $\Oc_1$ and $\Oc_2$ are spacelike separated regions,
then the elements of $\Group_L(\Oc_1)$ commute with those of 
$\Group_L(\Oc_2)$, as was explained above. This feature is 
passed on to the corresponding algebras $\Alg_L(\Oc_1)$ and 
$\Alg_L(\Oc_2)$, whose elements also commute with each other. 
Thus the net $\Oc \mapsto \Alg_L(\Oc)$ satisfies the 
condition of locality (Einstein causality). 

\medskip
As was also explained, each Poincar\'e transformation
$P \in \Pc$ determines an automomorphism 
$\alpha_P : \Group_L \rightarrow \Group_L$, 
fixed by the maps $S(F) \mapsto S(F_P)$, $F \in \Fc$.
It straightforwardly extends to the linear span of the 
elements of $\Group_L$. Now, 
$\| \, \cdot \, \|_P \doteq \| \, \alpha_P(\, \cdot \,) \, \|$
defines a C*-norm on this space. Since $\| \, \cdot \, \|$ is,
by definition, the (unique) supremum of its C*-norms, it implies
that $\| \, \alpha_P(\, \cdot \,) \, \| = \| \, \cdot \, \|$.
Thus $\alpha_P$ extends as an automorphism to the
C*-algebra $\Alg_L$. Moreover, the automorphisms $\alpha_P$, $P \in \Pc$,
act covariantly on the 
local algebras. This is a consequence of the fact that if
$F \in \Fc$ has support in $\Oc$,
then $F_P$ has support in $P \, \Oc$; it entails 
$$
\alpha_P\big(\Alg_L(\Oc)\big) = \Alg_L(P \, \Oc) \, , \quad
P \in \Pc \, , \ \Oc \subset \Mc \, .
$$ 

\medskip 
Thus the local net $\Oc \mapsto \Alg_L(\Oc)$ has the fundamental properties 
postulated by Haag and Kastler for any physically meaningful quantum field
theory on Minkowski space. In addition, these
authors require that the global algebra generated by a net 
should be primitive, \ie have some faithful irreducible representation.
This condition 
is motivated by their principle of physical equivalence according to 
which the states in any faithful 
representation should be weakly dense in 
the state space of any other representation. The algebra $\Alg_L$,
however, does not have this property since it has a non-trivial center
(containing for example the operators corresponding to the 
relative actions). This problem is solved by picking some 
irreducible
representation of $\Alg_L$ and taking the quotient with regard
to its kernel, being a primitive ideal of the algebra.  
In this way one abtains a primitive algebra, 
where as to yet unspecified physical data of the 
underlying quantum theory,
such as the field equation, the specific values of coupling 
constants and the mass are fixed.   
If the kernel of the representation is stable under the automorphic
action of the Poincar\'e
transformations, this quotient still defines a net with the preceding
desirable properties. So there arises the question of determining
such representations of physical interest. This is discussed
in the subsequent subsection.

\subsection{States of interest} 

Given a Lagrangean $L$, all possible states of $\Alg_L$
appear as elements of its dual space and determine corresponding 
representations by the GNS-construction. Pure states give rise to 
irreducible representations. 
In view of its manifold applications, the theory ought to 
describe states with a definite physical interpretation. 
These are primarily stable elementary systems and  
their excitations. On the other hand, the theory should reproduce 
quantitative results, obtained in the 
perturbative treatment of quantum field theory. 
As a matter of fact, these two issues are related. 

\medskip 
In order to exhibit this relation, let us consider the 
Epstein-Glaser method of renormalized perturbation theory \cite{EG};
it is based on power series expansions of correlation functions  
in terms of the scaled Planck constant. 
There one succeeds in constructing formal states on 
the linear span of operators, generating $\Alg_L$, 
\ie linear functionals which take values in the space of 
formal power series in $h$,
\[
\omega[h]=\sum h^k\omega_k \, .
\]
The functionals $\omega[h]$ satisfy in the sense of formal power series 
the positivity condition
\[
\omega[h](A^* A)=|\sum a_n \, h^n|^2 \, ,
\]
expressing the fact that it is a series with real coefficients whose 
lowest non-vanishing term is positive and of even order in $h$.
Their construction relies on local 
stability properties of states on the 
sub-algebra of bounded functions generated by 
smeared non-interacting fields, cf.\ the subsequent sections. 

\medskip
In general, one may not expect that these series 
converge. But in view of the empirical success 
of perturbative quantum field theory, one may hope that 
there exist states $\omega$ on $\Alg_L$ which determine corresponding 
formal states. In more detail, let 
\[
\omega_{\, h}(S(F_1)^{\sigma_1} \cdots S(F_n)^{\sigma_n}) \doteq 
\omega(S(hF_1)^{\sigma_1} \cdots S(hF_n)^{\sigma_n})  
\]
with $S(F_1), \dots , S(F_n) \in \Alg_L$
and $\sigma_i\in\{\pm1\}$, $i=1, \dots,n$. Then $\omega[h]$ should 
ideally 
describe the Taylor series at $h = 0$
corresponding to the function $h \mapsto \omega_h$. 
At present it is not known which precise conditions a 
state $\omega$ on $\Alg_L$
must satisfy in order to combine the  desired 
features. So we have to remain somewhat sketchy at this point. 
Based on insights gained in perturbation theory and 
basic properties of the algebras $\Alg_L$, we will 
indicate some promising conditions and call pure states satisfying
any one of them ``\primal states''. 

\medskip 
The most prominent 
examples of \primal states are vacuum states. 
They can be identified as follows.

\medskip \noindent 
\textbf{Definition:} A pure state $\omega_{\, 0}$ on $\Alg_L$ is said to be
a vacuum state if (i) $\omega_{\, 0} \scirc \alpha_P = \omega_{\, 0}$
and $P \mapsto \omega_{\, 0}(A_1 \alpha_P(A_2))$, $P \in \Pc$,
is continuous for all $A_1, A_2 \in \Alg_L$; \ (ii)
the Fourier transforms (in the sense of distributions) of \ 
$x \mapsto \omega_{\, 0}(A_1 \alpha_x(A_2))$, \ $x \in \RR^d$, \ have 
support in the forward lightcone $V_+$.

\medskip
It is a basic result in algebraic quantum field 
theory \cite[Sec.\ 4.2]{Ar} that these conditions imply that (i) the 
Poincar\'e transformations are unitarily implemented in the GNS
representation $\pi_0$ induced by $\omega_0$,  
(ii) the generators of the space-time translations 
(energy  and momentum) have joint
spectrum in~$V_+$, and (iii) $\omega_0$ is 
their ground state. Thus the kernel of $\pi_0$ is Poincar\'e 
invariant, 
so the net $\Oc \mapsto \Alg_L(\Oc) / \mbox{ker} \, \pi_0$ complies
with the Haag-Kastler postulates and $\Alg_L / \mbox{ker} \, \pi_0$
is a primitive C*-algebra. 

\medskip
As already mentioned, there 
exist theories of interest, such as the free massless field in 
$d=2$ dimensions, where such vacuum states do not exist. But one  
can relax the condition of Poincar\'e invariance of 
\primal states and also drop the assumption that 
the full Poincar\'e group is unitarily represented in the 
corresponding representations. In order to 
establish the required stability, it would suffice to exhibit 
\primal states, where 
only the space-time translations are unitarily implemented in
the corresponding representations, 
having generators with spectral properties as stated above. 
Or one may even be content with \primal states satisfying a microlocal 
version of the spectral condition \cite{BFK96}, which does not require the 
existence of generators. 

\medskip 
The condition that the kernels of the resulting 
irreducible representations $\pi$ are stable under Poincar\'e transformations
seems, however, to be inevitable on physical grounds. For, 
otherwise, the Poincar\'e group 
would not act on the resulting algebras $\Alg_L / \mbox{ker} \, \pi$;
it would be truly (not only spontaneously) broken. The condition
is satisfied by a \primal state $\omega$ 
if all Poincar\'e transformed states 
$\omega \scirc \alpha_P$ are locally normal 
with respect to each other, \ie if the restrictions of the 
resulting representations $\pi_P$ to any
given local algebra $\Alg_L(\Oc)$ are quasi-equivalent, $P \in \Pc$,
cf.~\cite[Def.\ III.2.2.15]{Haag92}. 
In the non-interacting case, these conditions are satisfied by 
so-called infra-vacuum states.

\medskip
Having chosen a \primal state $\omega$, one can 
determine the equation of
motion of the underlying field. It is encoded in the operators 
$S(\delta L(\phi_0))$, depending on 
the relative actions, which form a unitary group 
in the center of $\Alg_L$. Since the representation $\pi$, 
fixed by $\omega$, is irreducible, they are 
represented by phases, \ 
$\pi(S(\delta L(\phi_0)) \in \TT \, 1$. 
If the functions $u \mapsto \pi(S(\delta L(u \phi_0))$, $u \in \RR$, 
are continuous one can proceed to their derivatives.
In the absence of external sources,  
one then obtains the quantum analogue of the 
classical Euler-Lagrange equation,  
$$ \frac{d}{d u} \, \pi(S(\delta L(u \phi_0)) 
\Big|_{u = 0} = 0 \, , \quad  \phi_0 \in \Ec_0 \, .
$$  
It expresses 
the fact that the underlying quantum field corresponds to 
a saddle point of the action. In cases, where the field 
couples to an external (classical) source, this source manifests 
itself on the right hand side of this equality in the form of 
non-vanishing c-number contributions. 

\medskip
Given a Lagrangean $L$, it is, however, not clear 
whether the corresponding C*-algebra $\Alg_L$ has any 
\primal state in its dual space. This issue in the 
representation theory of C*-algebras  
is, from the present point of view, the  
remaining fundamental problem of constructive quantum field 
theory. Thinking for example of Lagrangeans with 
the common interaction potential
$\phi^4$, one expects on the basis of previous 
constructive results that \primal states (even vacua)  
can be found in $d =2$ and $d = 3$ dimensions \cite{GJ85}. Yet there
are also indications that in physical
spacetime $d=4$ and in higher dimensions such 
states do not exist \cite{Ai81,Fr81}. A proof 
of the presence or absence of physically acceptable \primal states 
within our algebraic setting would settle this matter, independently 
of any particular constructive scheme. 

\section{Non-interacting theories}
\setcounter{equation}{0}

As a first application of our approach, we discuss the case of
non-interacting scalar quantum fields in $d$ dimensions 
with masses $m \geq 0$. The corresponding classical Lagrangeans are
given by 
$$
x \mapsto  L_0(x)[\phi] \doteq 
1/2 \, (\partial_\mu \phi(x) \, \partial^\mu \phi(x) 
- m^2 \, \phi(x)^2 ) \, , \quad \phi \in \Ec \, .
$$
It is our goal to determine the algebraic properties of the quantum
fields in the corresponding algebras $\Alg_{L_0}$. 

\medskip
We consider functionals containing the sum of a 
linear term involving the underlying field and a constant functional. 
Let $K \doteq -(\square + m^2)$ be the Klein-Gordon operator, fixed by the 
Lagrangean~$L_0$, and let $\Delta_R$, $\Delta_A$ be the corresponding 
retarded and advanced propagators; their mean is the 
Dirac propagator $\Delta_D \doteq 1/2 \, (\Delta_A + \Delta_R$). 
These propagators define maps 
of the test function space $\Test(\Mc)$ into its dual space of distributions.
Making use of standard notation, the functionals have the form 
$$
F_f[\phi]  \doteq \phi(f) + 1/2 \langle f, \Delta_D \, f \rangle \, ,
\quad f \in \Test(\Mc) \, , \ \phi \in \Ec \, .
$$ 
Picking any field $\phi_0 \in \Ec_0 \simeq \Test(\RR^d)$, one obtains for  
test functions of the special form~$f = K \phi_0$ 
$$
F_f[\phi]  = \phi(K \phi_0) + 1/2 
\langle K \phi_0, \,  \Delta_D \, K \phi_0 \, \rangle 
= \phi(K \phi_0)  + 1/2 \langle \phi_0, K \phi_0 \, \rangle 
= \delta L_0(\phi_0)[\phi] \, .
$$
Thus for these special test functions the functionals coincide
with the relative actions.

\medskip 
We proceed now to the corresponding unitary operators 
$W(f) \doteq S(F_f)$ for arbitrary test functions
$f \in \Test(\Mc)$. As we shall see, these operators have the algebraic 
properties of exponentials of a free field (Weyl operators).
Given $f$, let $f = f_0 + K \phi_0$ be any decomposition  
with $f_0, \phi_0 \in \Test(\RR^d)$. Then 
\begin{align*}
F_f[\phi] & = \phi(f_0 + K \phi_0)  + 1/2 
\langle (f_0 + K \phi_0), \Delta_D \, (f_0 + K \phi_0) \rangle  \\
& =
\big((\phi + \phi_0)(f_0) + 1/2 \langle f_0, \Delta_D \, f_0 \rangle \big)
+ \big(\phi(K \phi_0)  + 1/2 \langle \phi_0, K \phi_0 \, \rangle \big) \\
& = F_{f_0}^{\, \phi_0}[\phi] + \delta L_0(\phi_0)[\phi] \, .
\end{align*}
Thus, making use of the dynamical relations between $S$-operators, we
obtain 
$$ 
W(f) = S(F_{f_0}^{\, \phi_0} + \delta L_0(\phi_0))
= S(F_{f_0}) \, S(\delta L_0(\phi_0)) =
W(f_0) W(K \phi_0) \, . 
$$
Given a second test function $g \in \Test(\Mc)$, we 
choose a decomposition of $f$ such that the support of 
$f_0$ is later than the support of $g$. 
That such a decomposition exists is a known fact; we 
briefly recall the argument  \cite{FNW81}. 
Let $C$ be a Cauchy surface lying in the
future of the support of $g$ and let $\chi$ be a smooth
function which is equal to $1$ in the future of $C$ and
tends to $0$ in its past at sufficiently small distance.
Making use of the equality $K \, \Delta_R f = f$, 
we define $f_0  \doteq K \chi \Delta_R f$ and 
$\phi_0 \doteq (1 - \chi) \Delta_R f$. 
Due to the support properties of $\Delta_R$, both expressions 
are test functions, $f = f_0 + K \phi_0$, 
and $f_0$ has its support in the future of $g$. 

\begin{figure}[h]
\centering
\includegraphics[width=0.6\textwidth]{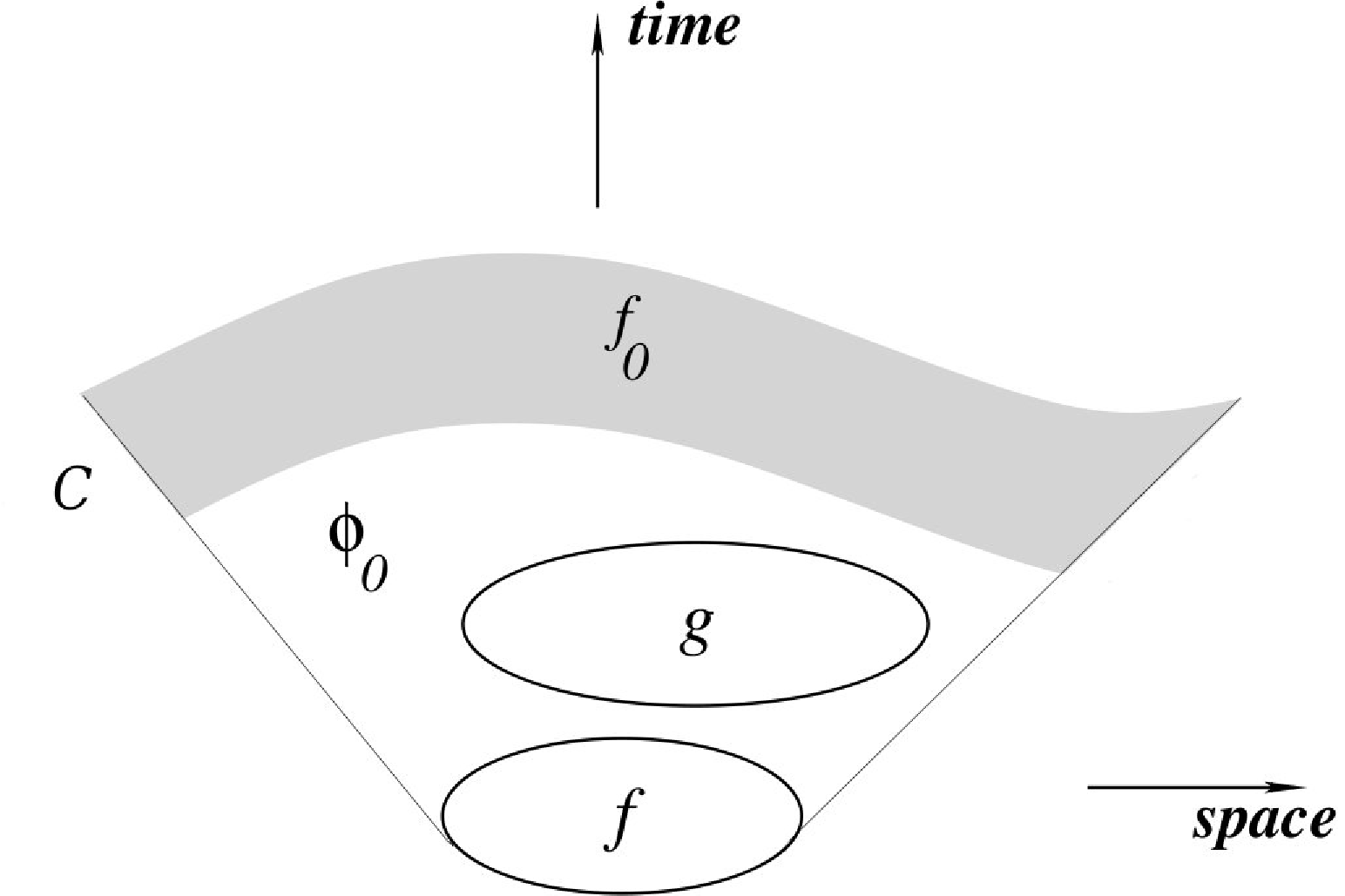}
\captionsetup{labelformat=empty}
\caption{Fig. \small \ Split $f = f_0 + K \phi_0$: $f_0$ has support 
in $C$ and $\phi_0$ in the entire region}
\end{figure}

\medskip
We exploit now these support properties of $f_0$, $g$. 
Since $S(F_{K \phi_0}) = S(\delta L_0(\phi_0))$, the defining properties of the
$S$-operators imply 
\begin{align*}
W(f) W (g) & = W(f_0) W(K \phi_0) W(g)
= S(F_{f_0}) S(F_{K \phi_0}) S(F_g) \\
& =  S(F_{f_0} + F_g) S(\delta L_0(\phi_0)) 
=  S(F_{f_0}^{\, \phi_0} + F_g^{\, \phi_0} +\delta L_0(\phi_0) ) \, .
\end{align*}

\vspace*{-1mm} \noindent 
Now, for $\phi \in \Ec$,  
\begin{align*}
& F_{f_0}^{\, \phi_0}[\phi] + F_g^{\, \phi_0}[\phi] +\delta L_0(\phi_0)[\phi] \\
& = (\phi + \phi_0)(f_0 + g) + 1/2 \langle f_0, \Delta_D f_0 \rangle
 + 1/2 \langle g, \Delta_D g \rangle
  + \phi(K \phi_0) + 1/2 \langle K \phi_0, \Delta_D K \phi_0 \rangle \\
& = \phi(f + g) + 1/2 \langle (f + g), \Delta_D (f + g) \rangle - 
\langle f_0, \Delta_D g \rangle \, .
\end{align*}
In view of the support properties of $\Delta_A$,  we have 
$ \supp f_0 \, \bigcap \, \supp \Delta_A \, g = \emptyset$. 
Hence 
$\langle f_0, \Delta_D g \rangle = 1/2 \, \langle f_0, \Delta_R g \rangle
= 1/2 \, \langle f_0, \Delta g \rangle$,
where $\Delta = \Delta_R - \Delta_A$ is the commutator function, 
which is a bi-solution of the Klein-Gordon equation. Thus 
$\langle K \phi_0, \Delta g \rangle = 0$, which altogether gives 
$\langle f_0, \Delta_D g \rangle = 1/2 \, \langle f, \Delta g \rangle$.
Since $1/2 \, \langle f, \Delta g \rangle$ is independent of 
$\phi \in \Ec$, it defines the constant functional
$F_{1/2 \, \langle f, \Delta g \rangle}$ on $\Ec$,
hence 
$$
F_{f_0}^{\, \phi_0}[\phi] + F_g^{\, \phi_0}[\phi] +\delta L_0(\phi_0)[\phi] 
= F_{f + g} - F_{1/2 \, \langle f, \Delta g \rangle} \, . 
$$
Plugging this relation into the preceding equality of $S$-operators, 
we arrive at 
$$
W(f) W(g) = W(f + g) \, e^{\, - i/2 \, \langle f, \Delta g \rangle} \, ,
\quad f,g \in \Test(\RR^d) \, .
$$
These are the Weyl relations of a free field;
since $\langle f, \Delta f \rangle = 0$, 
they imply in particular that $W(f)^{-1}=W(-f)$. 

\medskip 
By similar arguments one can also compute the product of 
Weyl operators with arbitrary $S$-operators $S(F)$, $F \in \Fc$,
cf.\ also the discussion in \cite[Sec.\ 4]{HoWa2}. \change
They give the equalities 
$$
W(f) S(F) = S(F_f + F^{\Delta_R f}) \, , \quad
S(F) W(f) = S(F_f + F^{\Delta_A f}) \, .
$$
They imply that the Weyl operators induce specific automorphisms of the 
$S$-operators, 
$$
W(f) S(F) W(f)^{-1} = S(F^{\Delta f}) \, , \quad f \in \Dc(\RR^d) \, . 
$$
The Weyl operators form an infinite dimensional
non-commutative subalgebra (Weyl algebra) 
of $\Alg_{L_0}$. In view of the preceding relations, 
involving also functionals $F$ 
depending on higher powers of the field $\phi$, 
it is apparent that the dynamical algebra $\Alg_{L_0}$
has an even more complex structure. As a matter of fact, one
can establish a natural correspondence between  local algebras in the  
interacting theories and 
subalgebras of~$\Alg_{L_0}$. This fact will be explained in the 
subsequent section. 

\medskip 
On the Weyl algebra there exist pure vacuum states.  
(In case of the massless free field in $d=2$ dimensions, there exist other 
\primal states.) 
Moreover, in the corresponding GNS-representations,
a perturbative expansion of all $S_h$-operators 
into formal power series in $h$ has been established.
It is not known, however, whether the full (unitary) $S_h$-operators 
can also be accommodated in these representations. This 
issue is reminiscent of non-commutative moment problems,
where an affirmative solution often requires an extension of 
the given Hilbert space. 
Since $\Alg_{L_0}$ is a C*-algebra, the pure states on its 
Weyl subalgebra can be extended to pure states on the full 
algebra by the Hahn-Banach theorem. So there remains the question 
of whether there exist extensions which are still \primal states.
 
\section{Interacting theories}
\setcounter{equation}{0}

Given a Lagrangean $L_0$ (which may differ from the Lagrangean of a 
free field), we 
study now the effect of perturbations of the dynamics, obtained  
by changes of the mass, the coupling constants, and the 
degree $N$ of the polynomial appearing in the 
interaction potential; the kinetic energy remains 
unaffected. The perturbed Lagrangean is denoted 
by $L_V \doteq L_0 + V$, where the perturbation $V$ has the form
$$
x \mapsto 
V(x)[\phi] \doteq  \sum_{n=0}^N \Delta \, g_n \, \phi^n(x)   
\, , \quad \phi \in \Ec \, , 
$$
with fixed variations  $\Delta g_n \in \RR$ of the coupling constants. 
Thus, by our constructive scheme, we are 
dealing now with two algebras, $\Alg_{L_0}$ and $\Alg_{L_V}$;
the corresponding $S$-operators will be denoted by 
$S_0$ and $S_V$, respectively. 

\medskip 
We want to show that the effect 
of the perturbations on the local algebraic properties of $\Alg_{L_V}$
can be fully described within the unperturbed  
algebra $\Alg_{L_0}$ \cite{IS78,BF00}. More precisely, given any  
bounded, causally closed region $\Oc \subset \Mc$ 
and any larger region $\widehat{\Oc}$, containing the closure 
of $\Oc$ in its interior, we will exhibit some subalgebra 
$\Alg_{L_0 + V(\chi)}(\Oc) \subset \Alg_{L_0}(\widehat{\Oc})$
which is isomorphic to $\Alg_{L_V}(\Oc)$. 
These subalgebras are not unique, but different
choices are related by inner automorphisms of 
$\Alg_{L_0}(\widehat{\Oc})$. Because of the latter fact,
there exists a homomorphic \change picture of the net 
$\Oc \mapsto \Alg_{L_V}(\Oc)$ within the algebra~$\Alg_{L_0}$
for any choice of interaction potential $V$. 

\medskip 
Turning to the proof of these assertions, we choose for 
given pair $\Oc \subset \widehat{\Oc}$ some smooth characteristic
function $\chi$ of $\Oc$ which has support in $\widehat{\Oc}$
and is equal to $1$ in an open neighbourhood of the
closure of $\Oc$. Integrating~$V$ 
with this test function, we obtain a functional 
$V(\chi) \in \Fc(\widehat{\Oc})$,
describing a perturbation which is localized in $\widehat{\Oc}$.
We also put \
$\delta L_{V(\chi)}(\phi_0) \doteq \delta L_0(\phi_0) + V(\chi)^{\phi_0} - V(\chi) $;
in view of the properties of $\chi$, this functional coincides with
$\delta L_{V}(\phi_0)$ for fields $\phi_0 \in \Ec_0$ having support in $\Oc$.
Adopting basic ideas of Bogolubov on the incorporation  
of interaction in quantum field theory, we define operators
(corresponding to the relative $S$-operators in the perturbative setting)
$$
\Bog_{\, V(\chi)}(F) \doteq S_0(V(\chi)\!)^{-1} S_0(F + V(\chi)\!) \, , 
\quad F \in \Fc \, .
$$
In view of the dynamical relations in $\Alg_{L_0}$, they
satisfy the equalities 
\begin{align*}
& \Bog_{\, V(\chi)}(F) \, \Bog_{\, V(\chi)}(\delta L_{V(\chi)}(\phi_0)\!) \\
&  = S_0(V(\chi))^{-1} S_0(F + V(\chi)\!) \ S_0(V(\chi)\!)^{-1} 
  S_0(\delta L_0(\phi_0) + V(\chi)^{\phi_0}) \\
& =  S_0(V(\chi)\!)^{-1} S_0(F + V(\chi)\!) \ S_0(\delta L_0(\phi_0)\!) =  
S_0(V(\chi)\!)^{-1} S_0(F^{\phi_0} + V(\chi)^{\phi_0} + \delta L_0(\phi_0)\!) \\
& = \Bog_{\, V(\chi)}(F^{\phi_0} + \delta L_{V(\chi)}(\phi_0)\!) \, .
\end{align*}
Moreover, for any functional $F_1$ having support in the future of
$F_2$ and any $F_3 \in \Fc$, the causal relations in $\Alg_{L_0}$ imply  
\begin{align*}
&  \Bog_{\, V(\chi)}(F_1 + F_3) \, \Bog_{V(\chi)}(F_3)^{-1}
  \Bog_{\, V(\chi)}(F_2 + F_3) \\
& =  S_0(V(\chi))^{-1} S_0(F_1 + (F_3 + V(\chi))\!) \ S_0(F_3 + V(\chi))^{-1}  
\ S_0(F_2 + (F_3 + V(\chi))\!) \\
& =  S_0(V(\chi))^{-1}  
S_0(F_1 + F_2 + F_3 + V(\chi)\!) \,  = \, \Bog_{\, V(\chi)}(F_1 +  F_2 + F_3)  \, .
\end{align*}

\medskip 
We restrict now the maps 
$\Bog_{\, V(\chi)} : \Fc \rightarrow \Alg_{L_0}$ to functionals $F$ 
having support in $\Oc$. Then 
$\delta L_{V(\chi)}(\phi_0)= \delta L_V(\phi_0)$ and 
the preceding two equations for the unitaries 
$\Bog_{\, V(\chi)}(F)$ coincide with the
defining relations of the perturbed $S$-operators $S_V(F)$, 
$F \in \Fc(\Oc)$. Let $\Group_{L_0 , \chi}(\Oc) \subset 
\Group_{L_0}(\widehat{\Oc})$ be 
the group generated by $\Bog_{\, V(\chi)}(F)$, $F \in \Fc(\Oc)$,
and let $\Alg_{L_0 + V(\chi)}(\Oc) \subset \Alg_{L_0}(\widehat{\Oc})$
be the corresponding C*-algebra. Identifying the 
operators $\Bog_{\, V(\chi)}(F)$ with $S_V(F)$, $F \in \Fc(\Oc)$, establishes 
an isomorphism $\beta_{\Oc, \, \chi}$ between 
the groups $\Group_{L_0 , \chi}(\Oc)$ and 
$\Group_{L_V}(\Oc)$. This isomorphism extends to congruent 
linear combinations of the group elements, which 
form norm dense subalgebras of 
$\Alg_{L_0 + V(\chi)}(\Oc)$ and $\Alg_{L_V}(\Oc)$, respectively. 
Denoting by $\| \, \cdot \, \|_0$ and $\| \, \cdot \, \|_V$
their original C*-norms, one obtains new C*-norms 
on these subalgebras, defined by  $\|\beta_{\Oc, \chi}( \, \cdot \, )\|_V$
and $\|\beta_{\Oc, \chi}^{-1}( \, \cdot \, )\|_0$, respectively. 
In view of the maximality of the original C*-norms, one has
on the respective subalgebras 
$$ \|\beta_{\Oc, \chi}( \, \cdot \, )\|_V \leq  \| \, \cdot \, \|_0 \, ,
\quad \|\beta_{\Oc, \chi}^{-1}( \, \cdot \, )\|_0 
\leq  \| \, \cdot \, \|_V \, .
$$
It implies $\|\beta_{\Oc, \chi}( \, \cdot \, )\|_V =  \| \, \cdot \, \|_0$
and $\|\beta_{\Oc, \chi}^{-1}( \, \cdot \, )\|_0 =  \| \, \cdot \, \|_V$, 
so the isomorphism $\beta_{\Oc, \chi}$ extends to the full 
C*-algebras. Moreover, the interpretation of the 
operators, based on the underlying classical functionals,  
does not change under its action. This establishes our first 
assertion, saying that the perturbed theory can locally be
described in terms of the unperturbed one. Interchanging the role
of $L_0$ and $L_V$, the converse is also true.

\medskip
Before we enter into the discussion of the dependence  
of our construction on the choice of $\chi$, let us briefly comment on 
its physical interpretation. To this end we return to  
the operators $\Bog_{\, V(\chi)}(F)$ for arbitrary $F \in \Fc$.
If $F$ has its support in the past of $ V(\chi)$, it follows 
from the causal relations that
$$
\Bog_{\, V(\chi)}(F) = S_0(V(\chi))^{-1} 
S_0(F + V(\chi)) = S_0(F) \, .
$$
Similarly, if $F$ has its support in the future of $ V(\chi)$, one gets 
$$
\Bog_{\, V(\chi)}(F) = S_0(V(\chi))^{-1} S_0(F + V(\chi)) = 
S_0(V(\chi))^{-1} S_0(F) \, S_0(V(\chi))
\, . 
$$ 
Thus the map $\Bog_{\, V(\chi)}$ describes a perturbation 
of the original theory 
in the region $\widehat{\Oc}$, leaving it
unaffected in the past and spacelike complement of $\widehat{\Oc}$. 
Due to the interaction, the unperturbed theory is affected, however,
in $\widehat{\Oc}$ and its causal future. 
Alluding to a dynamical picture, one may think of the  
unperturbed system as coming in from large negative times and, 
eventually, reaching the localized potential 
in $\widehat{\Oc}$. The overall effects of this perturbation 
become visible in the future of $\widehat{\Oc}$ and can be reinterpreted 
in terms of the original (similarity transformed) theory.
Thus the similarity transformation \ $\ad{S(V(\chi))^{-1}}$ 
has the meaning of a scattering automorphism. 

\medskip 
Depending on the choice of $\chi$, the inclusion
$\Alg_{L_0 +V(\chi)}(\Oc) \subset\Alg_{L_0}(\widehat{\Oc})$ 
holds for regions $\widehat{\Oc}$ which are arbitrarily
close to the given region $\Oc$. And for any such 
choice of~$\chi$, the algebra $\Alg_{L_0 + V(\chi)}(\Oc)$ is isomorphic 
to $\Alg_{L_V}(\Oc)$. But this does \textit{not} imply, that
the Haag-Kastler nets $\Oc \mapsto \Alg_{L_0}(\Oc)$
and $\Oc \mapsto \Alg_{L_V}(\Oc)$ are isomorphic.
That is, in general there does not exist a  
global isomorphism between $\Alg_{L_0}$ and $\Alg_{L_V}$, 
which also maps the local subalgebras onto each other 
for all regions $\Oc \subset \Mc$. 
At best, one can hope that algebras corresponding
to causally closed regions, having their base in a 
common Cauchy surface, can be identified in the two 
theories.
This identification of local
algebras does not persist, however, at other times due to the 
differing interactions. Nevertheless, the
perturbed net $\Oc \mapsto \Alg_{L_V}(\Oc)$ 
can be accommodated in the unperturbed global algebra~$\Alg_{L_0}$,
as we shall show next. 

\medskip
Given a region $\Oc \subset \Mc$, we pick any two test functions
$\chi_1, \chi_2 \in \Dc(\RR^d)$, which, both, are equal to $1$ on 
$\Oc$ and have support in a region
$\widehat{\Oc} \supset \Oc$. In a first step we show that the two
subalgebras $\Alg_{L_0 +V(\chi_1)}(\Oc), \ 
\Alg_{L_0 +V(\chi_2)}(\Oc) \subset \Alg_{L_0}(\widehat{\Oc})$ are related by an 
inner automorphism of $\Alg_{L_0}(\widehat{\Oc})$. 
To this end we choose a decomposition of the difference 
$\chi_2 - \chi_1 = \chi_+ + \chi_-$, such that 
the support of $\chi_+$ does not intersect the past of $\Oc$ 
and that of $ \chi_-$ its future. 
(This is possible since the regions considered here are causally 
closed, cf.\ the figure below.)

\begin{figure}[h]
\centering
\hspace*{13mm} \includegraphics[width=0.6\textwidth]{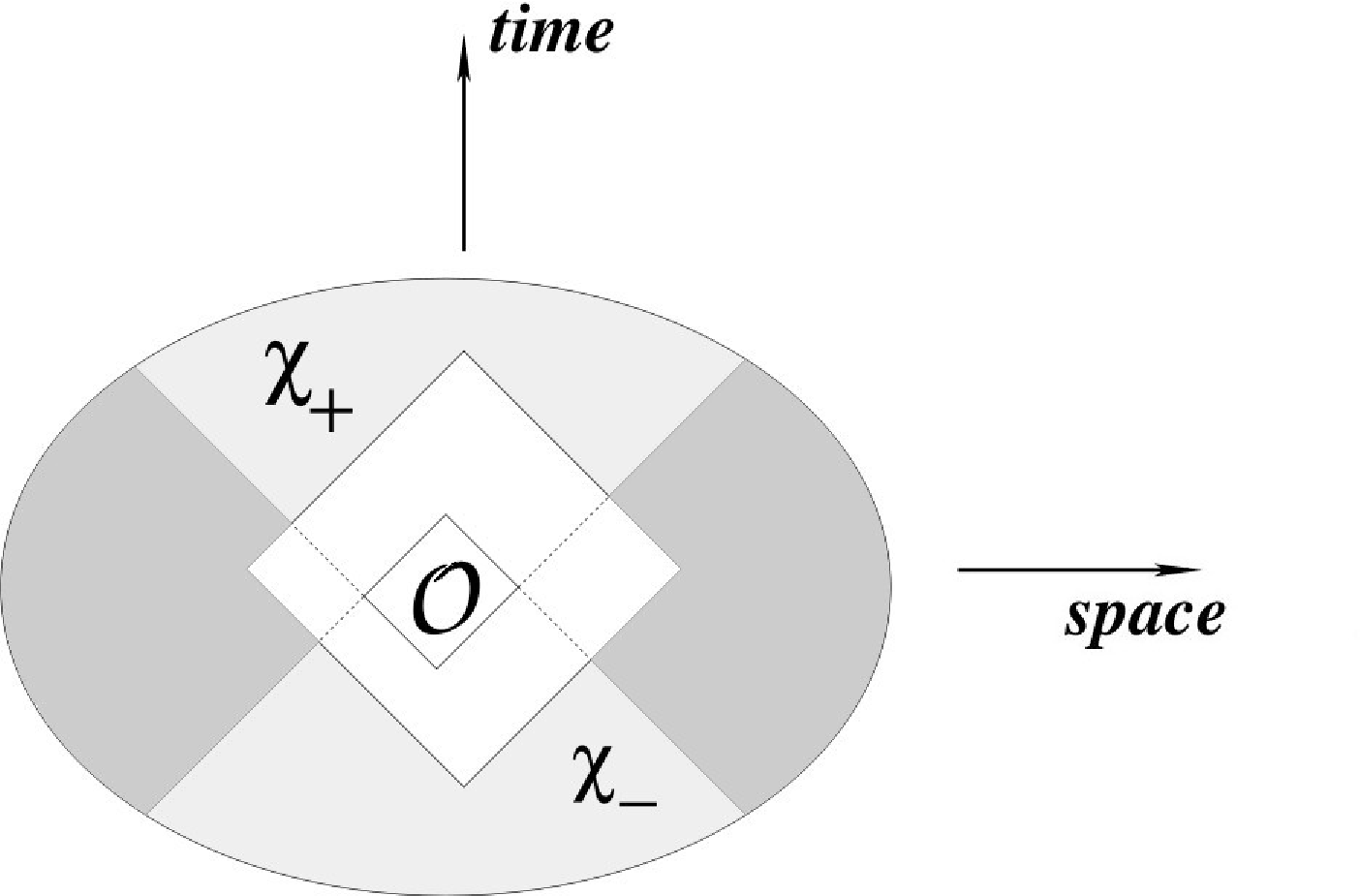}
\captionsetup{labelformat=empty}
\caption{\small Fig. \ The supports of $\chi_+, \, \chi_-$ are    
empty in the white and overlapping in the dark regions}
\end{figure}

Picking any $F \in \Fc$ which has support in $\Oc$
and making use of the fact that
$\chi_+$ has its support in the future of $\Oc$, we get from the 
causal relation for $S$-operators 
$$
S_0(V(\chi_2) + F) = 
S_0(V(\chi_+) + V(\chi_1 + \chi_-)) \, S_0(V(\chi_1 + \chi_-))^{-1}
\, S_0(V(\chi_1 + \chi_-) + F) \, .
$$
Similarly, since $\chi_-$ has its support in the past of $\Oc$,
we obtain 
$$
S_0(V(\chi_1 + \chi_-) + F)) =
S_0(F + V(\chi_1)) \, S_0(V(\chi_1))^{-1} \, S_0(V(\chi_1 + \chi_-))  \, .
$$
Combining these relations and inserting the result 
into the corresponding Bogolubov operators gives
$$
\Bog_{\, V(\chi_{\, 2})}(F) = \Bog_{\, V(\chi_1)}(V(\chi_-))^{-1}
 \Bog_{\, V(\chi_1)}(F) \ \Bog_{\, V(\chi_1)}(V(\chi_-)) \in 
 \Alg_{L_0}(\widehat{\Oc})  \, .
$$
It implies that the two isomorphisms $\beta_{\Oc, \chi_1}^{-1}, 
\beta_{\Oc, \chi_{\, 2}}^{-1}$, mapping 
$\Alg_{L_V}(\Oc)$ into corresponding subalgebras of $\Alg_{L_0}$, 
are related by 
$$ 
\ad{\big( \Bog_{\, V(\chi_1)}(V(\chi_-))^{-1} \big) }
 \scirc \, \beta_{\Oc, \chi_1}^{-1} = \beta_{\Oc, \chi_2}^{-1} \, .
$$
It shows that, disregarding inner automorphisms, 
the embeddings of the algebra $\Alg_{L_V}(\Oc)$
into $\Alg_{L_0}$ do not depend on the choice of
test functions $\chi$ whose restrictions to $\Oc$
are equal to $1$. 

\medskip 
For the proof that the entire C*-algebra $\Alg_{L_V}$ 
can consistently be accommodated in~$\Alg_{L_0}$, one has to adjust   
the embeddings of the local algebras $\Alg_{L_V}(\Oc)$,
$\Oc \subset \Mc$, into $\Alg_{L_0}$; \change we adopt here 
arguments given in \cite[Prop.~3.1]{HoWa1}.
Let $\Oc_n \subset \widehat{\Oc}_n \subset \Oc_{n+1}$, $n\in \NN$,  
be a sequence of bounded regions, such that the closure of 
$\Oc_n$ is contained in the 
interior of $\widehat{\Oc}_n$  and $\lim_n \, \Oc_n \nearrow \Mc$.  
Furthermore, let $\chi_n \in \Test(\RR^d)$ be a sequence 
such that $\chi_n \upharpoonright \Oc_n = 1$ and
$\, \supp \chi_n \subset \widehat{\Oc_n}$, $n \in \NN$. 
As we shall see, the corresponding isomorphisms \ 
$\beta_{\Oc_n, \chi_n}$, mapping $\Alg_{L_V}(\Oc_n)$ into 
$\Alg_{L_0}(\widehat{\Oc_n})$,  
can be transformed into a coherent sequence of 
isomorphisms $\gamma_{\Oc_n}$; their restrictions, respectively 
ranges, satisfy 
$\gamma_{\Oc_{n+1}} \upharpoonright \Alg_{L_V}(\Oc_n) = \gamma_{\Oc_n}$
and $\gamma_{\Oc_n}(\Alg_{L_V}(\Oc_n)) \subset \Alg_{L_0}(\widehat{\Oc_{n+1}})$, 
$n \in \NN$.
Since $\Alg_{L_V}$ is the C*-inductive limit of its local subalgebras,
this implies that there exist in the sense of pointwise norm convergence 
the limits
\[
\gamma(A_{L_V}) \ \doteq \ \lim_n \gamma_n(A_{L_V}) \, , \quad 
A_{L_V} \in \Alg_{L_V} \, .
\]
The limit $\gamma :  \Alg_{L_V} \rightarrow \Alg_{L_0}$
is a monomorphism (injective homomorphism),
\change mapping the local net $\Oc \mapsto \Alg_{L_V}(\Oc)$
in the perturbed theory, into a net of subalgebras of $\Alg_{L_0}$.
Moreover, for the given sequence of regions,
the algebras $\Alg_{L_V}(\Oc_n)$ are mapped into 
$\Alg_{L_0}(\widehat{\Oc_n})$,
$n \in \NN$. As already mentioned, this does not imply that a similar (fuzzy) 
identification of local subalgebras holds 
for all regions $\Oc \subset \Mc$. 

\medskip 
Turning to the construction of the sequence $\gamma_{\Oc_n}$, $n \in \NN$, 
we proceed from the isomorphism $\beta_{\Oc_n, \chi_n}$ for given $n$.
As we have shown, replacing in this 
isomorphism the underlying test function $\chi_n$ 
by $\chi_{n+1}$ amounts to transforming it by some 
inner automorphism of $\Alg_{L_0}(\widehat{\Oc}_{n+1})$. For the sake of brevity, 
these inner automorphism are denoted by $\ad{V_{n+1,n}}$, where 
$V_{n+1,n} \in \Alg_{L_0}(\widehat{\Oc}_{n+1})$, \
\ie 
\[\ad{V_{n+1,n}} \, \scirc \, \beta_{\Oc_n, \chi_n} =
\beta_{\Oc_n, \chi_{n+1}} \, , \quad n \in \NN \, .
\]
By construction, we also have \ 
$\beta_{\Oc_{n+1}, \chi_{n+1}} \upharpoonright \Alg_{L_V}(\Oc_n) =
\beta_{\Oc_n, \chi_{n+1}}$. We define now 
\[
\gamma_{\Oc_n} \doteq \ad{\big(V_{n,n-1} \cdots V_{2,1} \big)^{-1}} 
\, \scirc \ \beta_{\Oc_n, \chi_n} \, , \quad n \in \NN \, ,
\]
where $V_{1,0} \doteq 1$. The preceding preparations imply that 
$\gamma_{\Oc_n}(\Alg_{L_V}(\Oc_n) \subset \Alg_{L_0}(\widehat{\Oc}_n)$.
Moreover, we obtain for $n \in \NN$
\begin{align*}
& \gamma_{\Oc_{n+1}} \upharpoonright \Alg_{L_V}(\Oc_n)
=  \ad{\big(V_{n,n-1} \cdots V_{2,1} \big)^{-1}} \ \scirc \   
 \ad{V_{n+1,n}^{-1}} \ \scirc \ 
\beta_{\Oc_{n+1}, \chi_{n+1}} \upharpoonright \Alg_{L_V}(\Oc_n) \\
& = \ad{\big(V_{n,n-1} \cdots V_{2,1} \big)^{-1}} \ \scirc \ 
\ad{V_{n+1,n}^{-1}} \ \scirc \ 
\beta_{\Oc_n, \chi_{n+1}} \upharpoonright \Alg_{L_V}(\Oc_n) \\
& = \ad{\big(V_{n,n-1} \cdots V_{2,1} \big)^{-1}} \ \scirc \ 
\beta_{\Oc_n, \chi_n} \upharpoonright \Alg_{L_V}(\Oc_n)
\ = \ \gamma_{\Oc_n} \, ,
\end{align*}
establishing the existence of monomorphisms \change with the desired 
properties. 

\medskip 
This completes 
our proof that for any choice of Lagrangean $L$, 
the corresponding local nets $\Oc \mapsto \Alg_L(\Oc)$ can \change 
be embedded into a fixed dynamical C*-algebra $\Alg_{L_0}$. 
For given $\Alg_{L_0}$, the nets induced by different theories 
differ by the assignment of 
subalgebras of $\Alg_{L_0}$ to a given 
spacetime region. The global algebra 
$\Alg_{L_0}$ does not contain any specific dynamical information by 
itself. 

\medskip 
The existence of a global algebra containing the local nets of
a large family of theories has also been established 
in the axiomatic framework of algebraic quantum field theory. For 
mathematical convenience, one works there with local von Neumann algebras.
Whenever a theory has the so-called split property 
\cite{DoLo}, which amounts to 
restrictions on the number of degrees of freedom of the 
theory in finite volumes of phase space \cite{Haag92}, 
the corresponding nets can 
isomorphically be embedded into a unique global 
C*-algebra, the ``proper sequential type I$_\infty$ funnel'', 
invented by Takesaki \cite{Ta}. In the present approach we work 
in the setting of C*-algebras. The passage to local von Neumann 
algebras would require
to select some states of interest, inducing a weak topology on
the algebra. Apart from this technical point, the existence of a universal 
global algebra does not come as a surprise in the present approach. 

\medskip 
It might perhaps be more surprising that, using
the interaction picture, 
free and interacting theories can be placed into one and the same 
algebra. This seems to contradict Haag's theorem, 
which says that the combination of free and interacting  
Hamiltonians into a single interaction operator is impossible  
in quantum field theory. This obstruction is avoided in 
the present approach by dealing with a local version of the 
interaction picture, based on perturbations of the dynamics in 
finite spacetime regions.
In this way large volume  as well as ultraviolet 
singularities are avoided and a consistent
theory results.

\section{Summary and outlook}
\setcounter{equation}{0}

We have constructed in this article for any given Lagrangean of a 
real scalar field in Minkowski space a corresponding dynamical 
C*-algebra. 
In order not to obscure the underlying ideas, we have 
restricted our attention to interactions given by 
polynomials of the field; but our construction works for 
quite arbitrary interaction potentials. 
The novel input in our construction is an integrated version
of the Schwinger-Dyson equation in terms of 
unitary $S$-operators, 
containing information encoded in the field equations. 
The second ingredient are causal factorization rules for 
these operators. Both features 
have been established in the perturbative setting of quantum
field theory, based on formal power series in Planck's constant. 
The resulting equalities, established in the latter approach, were 
taken as input in our construction of the C*-algebras. 
Let us emphasize that these algebras exist for any number 
of spacetime dimensions, any degree of the interaction 
polynomial and any choice of coupling constants, irrespective of 
their signs. It is the existence of  
states of physical interest on these algebras, named \primal
states in the present article, which is expected to depend on the
choice of these parameters. Thus the question of  
whether there are such states in a particular
model has been traced back to a problem in the representation
theory of C*-algebras. 

\bigskip
The C*-algebras, obtained by our construction,  
have a quite non-trivial structure. Without having to impose from the outset 
any quantization rules, they are intrinsically non-commutative.
Taking as input a non-interacting Lagrangean, the exponentials of
the underlying field were shown to satisfy the 
Weyl relations, and there exist unitary operators in the algebra  
which can be interpreted as time ordered exponentials of its 
normal ordered powers, integrated with test functions. 
More importantly, we have shown that the resulting C*-algebra is 
universal in the sense that (up to isomorphisms) it coincides 
with the C*-algebras obtained from Lagrangeans describing 
arbitrary local interactions. 

\medskip
The latter result is in accord with the known fact that the 
physical information encoded in a theory is not contained in the
global algebra, generated by the fields, but in the  
assignment of its subalgebras to spacetime regions, \ie  
in the corresponding net of local algebras.  This basic 
insight found its expression in the postulates of  
algebraic quantum field theory, formulated by Haag and Kastler 
60~years ago. As a matter of fact, based on findings 
in his work on collision theory, Haag was convinced that
one can recover from a given net the entire physical 
content of the underlying theory. But his program to specify
a specific theory through properties of the corresponding 
net of local algebras remained unfinished \cite{HaOj}. 

\medskip
We have shown that the dynamical 
C*-algebras, constructed here, comply with all Haag-Kastler postulates.
Yet, in contrast to the ideas of Haag, we have used a bottom-up 
approach, where the operators, generating the algebra, are labelled from the 
outset by physical quantities, such as ``field'', 
``interaction potential'', ``relative action'' \etc. 
In the spirit of Bohr, we regard these labels as 
notions in the framework of classical 
field theory, which are merely used to describe 
which kind of objects in the quantum world we have in mind. 
There is no \textit{a priori}
quantization rule for them. The  realization 
of the corresponding 
operators in the mathematical setting of quantum theory is 
fixed by the notion of causality, involving time ordering; 
the actual results depend on the chosen Lagrangean. 
These observations solve the longstanding problem of incorporating
a dynamical principle into the Haag-Kastler framework. 

\medskip
The present results seem  to suggest, however, a change of paradigm in
the interpretation of the Haag-Kastler framework. Originally, it was 
proposed to interpret the selfadjoint elements of the 
local algebras as observables.  Yet such an interpretation does not 
fit well with our construction. To explain this point, 
consider for example the unitary $S$-operator, labelled by 
a localized interaction potential. It should \textit{not} be 
interpreted as (function of) a quantum observable, describing
the potential in the sense of the statistical interpretation of 
quantum physics. In view of the time ordering involved
in the construction of $S$-operators, such an interpretation
would be meaningful if measurements could be performed 
instantaneously; only then time ordered and unordered operators 
do coincide. Yet, restricting observables to a Cauchy surface makes them 
in general ill-defined and requires some smoothing in time,
surpressing their infinite fluctuations. In more physical terms,
reliable measurements require time, and this fact is taken
into account in the time-ordered $S$-operators. 
Unfortunately, this step blurs in general the interpretation of the 
envisaged observables due to perturbations 
caused by the interaction. 

\medskip 
As a way out of this conceptual dilemma, we propose to interpret the  
unitary operators as operations, describing the impact 
of measurements of the conceived \mbox{observables} on 
quantum states in the given 
spacetime regions. According to this view, a net of local algebras 
subsumes these operations. It is of interest in this context
that in \primal states, complying with the split property 
mentioned above, one can recover from the operations the
standard interpretation of states in terms of ``primitive 
observables'', which have a consistent statistical interpretation
in accordance with basic principles of quantum physics \cite{BuSt}. 

\medskip 
In the context of the  present family of models, there remain
some issues of physical interest which were not discussed in this article. 
First, besides the observables considered here, there exist other 
prominent examples, such as the kinetic energy, the stress energy 
tensor, the full Lagrangean \etc. They have in common that they 
involve derivatives of the underlying classical field. The resulting 
functionals therefore require some qualifications in applications of 
the causal factorization relation, which go beyond 
formal perturbation theory. One has to ensure that the pertubations 
admitted in these relations are compatible with the causal structure 
of Minkowski space. In case of local functions of the field, considered here, 
this condition is automatically satisfied and was therefore not mentioned. 
In the general case, the constraints on the functionals 
can be expressed by conditions on the 
correspondingly perturbed Euler-Lagrange derivatives.
The dynamical C*-algebras then  exist for the 
enlarged set of observables. Second, in this enlarged framework one 
can study the behaviour of 
symmetry transformations, such as the space-time translations, 
under perturbations of the dynamics. They determine cocycles in the
C*-algebra. It is an interesting, but more difficult problem, whether 
there exists also an analogue of Noether's theorem in the present 
setting, establishing the existence of currents inducing these 
symmetry transformations. 
We will return to these topics in a future publication.

\medskip 
The remaining major problem concerns the existence of \primal states.
At present, there are two strategies visible towards its solution. The 
first one is based on perturbation theory, where one can try to prove 
convergence of the formal power series of the $S$-operators. This works
indeed in some exceptional cases, as for example for 
quadratic functionals of the field \cite{Bel,Rui} 
and in models in two dimensional Minkowski space, such as the 
$\phi^4$  and Sine-Gordon theories, cf.\  \cite{Wreszinski} 
and \cite{BaRe}. The 
second one is based on the development of criteria in the C*-algebraic 
setting, implying the existence of \primal states. Such a criterion  
was proposed by Doplicher in \cite{Do}. It works in case of 
C*-dynamical systems; but it is not applicable here, 
since the present algebras are lacking the continuity properties 
under the action of symmetry transformations, required in the criterion. 
So some progress is needed on this algebraic side. 

\medskip
Further steps in this program are the extension of our construction 
to theories with Lagrangeans, involving several Bose and Fermi fields, 
vector fields and, ultimately, also local gauge groups. We believe that 
in spite of the indefinite metric entering in the conventional perturbative
treatment of such theories, they can be transferred into a C*-algebraic 
setting. Limitations on a meaningful physical interpretation 
of the underlying fields will manifest themselves in the absence
of \primal states. These will in general only exist on 
(gauge invariant) 
subalgebras. In the construction of the corresponding C*-algebras, 
there appear several new problems, such as the 
formulation of Ward identities, the characterization of anomalies and 
the description of BRS-transformations. They all have to be cast 
into a proper C*-algebraic form, in analogy to the Schwinger-Dyson 
equations used in the present article. The present C*-algebraic 
approach can also be extended to interacting quantum field theories 
on curved spacetimes. There the generally covariant 
locality principle \cite{BFV03}, which is basic in the treatment of 
these theories, finds its most natural formulation. 
Even though we are only at the very beginning of these developments, 
we are confident that the present novel approach will 
contribute to the long hoped-for 
mathematical consolidation of quantum field theory.

\appendix
\section{Appendix}
\setcounter{equation}{0}

In this appendix we show that the dynamical relation, taken as 
input in the present C*-algebraic setting, can be established in 
the sense of formal power series in renormalized perturbation 
theory. We also give a brief account of  the 
mathematical framework underlying our arguments. 

\medskip 
There exist several rigorous approaches to the perturbative treatment 
of quantum field theories. The framework used here, based on classical
off-shell fields, is related to the path integral formulation
of quantum field theory and to deformation quantization. 
The former approach is successful for perturbative computations in 
Euclidean space, based on expansions in terms of Planck's
constant; but the interpretation of the results in Minkowski space
requires analytic continuations. The latter approach 
works in physical spacetime and generates power series in 
Planck's constant of the quantized fields. 
But it is usually restricted to on-shell fields of the classical 
theory, which hampers the comparison of different theories. 

\medskip
We therefore work in the framework of algebraic perturbation theory,
based on Minkowski space, which is conceptually  
very close to algebraic quantum field theory, 
cf.\ for example \cite{Rej,FR16}. 
The starting point are functionals $F$ on the underlying 
classical configuration space. In the case of a real scalar field this is 
the space of smooth real functions $\Ec$ on Minkowski space 
$\Mc$; the functionals map arbitrary field configurations to complex numbers. 
The vector space
of functionals is equipped with three associative products:
the pointwise product, \tpunkt, the non-commutative 
product~\tstern, which corresponds to the operator
product in quantum theory (used in the main text), and 
the time ordered product \tzeit.

\medskip 
For functionals of the form $e^{i\phi(f)}$ where $f \in \Dc(\RR^d)$,
$\phi \in \Ec$, these products are defined by
\begin{align*}
e^{i\phi(f)} \, \mpunkt \ \, e^{i\phi(g)} &  \doteq e^{i\phi(f+g)} \\
e^{i\phi(f)}  \, \mstern \ e^{i\phi(g)} & \doteq 
e^{i\phi(f+g)}e^{-i/2\langle f,\Delta g\rangle} \\
e^{i\phi(f)}  \, \mzeit \ e^{i\phi(g)} & \doteq e^{i\phi(f+g)} 
e^{-i\langle f, \Delta_D g\rangle}
\end{align*}  
with the commutator function $\Delta$ and the Dirac propagator 
$\Delta_D=(1/2) (\Delta_R+\Delta_A)$ as in Sec.~4.
For later use we also define the time ordered exponential of the field, 
characterized by the functional equation
$$
e_T^{i\phi(f)} \, \mzeit \ \, e_T^{i\phi(g)} = e_T^{i\phi(f+g)} 
$$
and given by
$$
e_T^{i\phi(f)} \doteq e^{i\phi(f)} \, e^{-i/2\langle f,\Delta_Df\rangle}\ .
$$
In a similar manner  one can define \tstern -exponentials of the field 
which coincide with the \tpunkt -exponentials because of the 
antisymmetry of the  commutator function $\Delta$.

\medskip 
Due to the singularities of the propagators, the 
\tstern -product and the \tzeit -product cannot directly be extended 
to all functionals of interest here. In particular, they are 
undefined for 
non-linear local functionals occuring in typical interaction Lagrangeans, 
such as \ $V(f)[\phi]=\int \! dx \, f(x) \phi^4(x)$. 
The ill-posed problem 
of defining the \mbox{\tstern -product} can be circumvented,
however, by {\it normal ordering}. It is defined as follows.

\medskip 
Let $F$ be functionals of the form 
\begin{equation} \label{eA.1}
F[\phi]=\sum_{k=0}^n \, 
\int \! dx_1 \cdots dx_n \  f_k(x_1, \dots ,x_k) \,
\phi(x_1)\cdots\phi(x_k) \, , \quad \phi \in \Ec \, ,
\end{equation}
where $f_k$, $k=0, \dots , n$,  are test functions. 
Their normal ordering is defined by a linear invertible map
\ $F \mapsto \no{F}$ \ with generating function 
\[
\no{e^{i\phi(f)}} \  \doteq \ e^{i\phi(f)} \, e^{(1/2) \, ||f||_1^2} \, .
\]
Here $||f||_1$ is the familiar single-particle norm in $d$ 
spacetime dimensions for particles with mass \ $m \geq 0$; \ it is given by 
\[ ||f||_1^2=
(2\pi)^{-(d-1)} \int \! dp \, \theta(p_0) \delta(p^2 - m^2) \, 
| \widetilde{f}(p) |^2 \, ,
\]
where $\widetilde{f}$ denotes the Fourier transform of $f$. 
(If $m=0$, his expression is in general undefined in $d=2$ dimensions; 
yet there exist other suitable Hilbert norms.) \ One 
thus obtains the relation
\[
\no{e^{i\phi(f)}} \mstern \no{e^{i\phi(g)}} \ = \ 
\no{e^{i\phi(f+g)}} \, e^{\langle f,g\rangle_1} \ \doteq \ 
 :e^{i\phi(f)} \, \mnormal \ \, e^{i\phi(g)}: 
\]
with the scalar product $\langle \cdot \, , \, \cdot \rangle_1$ in 
the 1-particle space. The new product \tnormal \ is called
Wick-star product. 

\medskip
The Wick-star-product can be extended from functionals of the form  
\eqref{eA.1} to so-called microcausal functionals \cite{BF00}.
These are functionals as in equation \eqref{eA.1},
where the test functions  $f_k$, $k \in \NN$, are replaced by the 
larger set of distributions, whose wave front sets $W \! F(f_k)$
\ (cf.~\cite{Hoer})
are restricted by the following condition,  
\[
W \! F(f_k)\cap \big((\Mc^k,V_+^k)\cup(\Mc^k,V_-^k)\big)=\emptyset \, ,
\quad k \in \NN \ ;
\]
here $V_\pm$ are the closed forward, respectively backward, lightcones. 
This condition includes in particular translationally invariant
distributions, multiplied with test functions. For the corresponding
functionals, the product
was shown to be well defined in \cite[Thm.\ 0]{EG}. 
The algebra of regular functionals with respect to the product 
$\tstern$ \ can then be extended to normal ordered microcausal functionals, 
denoted by $\no{F}$. They satisfy
\begin{align*}
\no{F} + c \! \no{G}  =\no{F & + c \, G} \, , \ \ c \in \CC \\ 
\no{F}\mstern\no{G} & =\no{F \ \mnormal \ G} \, .
\end{align*}

An analogous extension of the time ordered product 
would be more complicated since the Dirac propagator is not a solution of 
the Klein Gordon equation. Using the normal ordering procedure 
amounts to transforming the time ordered product in a manner
such that the Dirac propagator is replaced by the Feynman propagator. 
But also after normal ordering, the time ordered 
product is not always well defined on microcausal functionals. 
In order to cure this defect some further steps are necessary.

\medskip 
One first observes that the $n$-fold products of local functionals  
are well defined whenever these functionals have disjoint supports.
These products admit extensions to local functionals with 
arbitrary support in compact regions. But these extensions are not unique;
they can be classified and parametrized by renormalization conditions, 
related to those used in other perturbative schemes \cite{EG}. 
Proceeding to the resulting multilinear products on the space 
of normal ordered local functionals, one finds \cite{FR13} that 
these can be understood as iterated binary products $\tzeit$ \ on a 
subspace of normal ordered microcausal functionals.
That space also contains the normal ordered local functionals and 
their time ordered products. The 
$\tzeit$ \ product is commutative and associative on this space. 

\medskip
One of the renormalization conditions which fix the time ordered 
products is the Schwinger-Dyson
equation. It characterizes the 
field equation of the interacting quantum fields and  has the form, 
$K$ being the Klein-Gordon operator, 
\begin{equation} \label{eA.2}  
\no{F} \mzeit \ \, {\phi(K\phi_0)} \, = \, \no{F} \mstern \ \, 
\phi(K\phi_0) + i\no{ \epsilon F(\phi_0) } \ .
\end{equation}
Here the functional derivative $\epsilon F$ of $F$ in the direction 
of $\phi_0$ enters, cf.~Sec.~2.
For regular functionals $F$, equation \eqref{eA.2} is a direct consequence 
of the definition of the various products considered here. The equation
still holds for microcausal functionals after the extension procedure, 
described above.

\medskip 
The Schwinger-Dyson equation has the form of a differential equation. 
We can integrate it in the following way. Let
\[
\no{F_{\lambda}}  \ \doteq \
\no{F^{\lambda \phi_{ \ 0}}} \ \mzeit \ e_T^{i\phi(K\lambda\phi_0)} \,
e^{i (\lambda^2/2) \,  \langle \phi_0, K \phi_0 \rangle}
\]
where $F^{\lambda \phi_0}[\phi] \doteq F[\phi+ \lambda \phi_0]$.  Then
\begin{align*}
{\textstyle \frac{d}{d\lambda}} \no{F_\lambda} & 
\, = \ \no{\epsilon F^{\lambda \phi_0}(\phi_0)} 
\ \mzeit \  e_T^{i\phi(K\lambda\phi_0)} 
e^{i (\lambda^2/2) \,  \langle \phi_0, K \phi_0 \rangle} \,             
+i \no{F_\lambda} \, \mzeit \ 
(\phi(K\phi_0)+\lambda\langle\phi_0,K\phi_0\rangle) \\
& \,  = \, \epsilon \no{F_\lambda}(\phi_0) + \, 
i \no{F_\lambda} \mzeit \ \, \phi(K\phi_0) \, 
= \, i \no{F_\lambda} \mstern \ \,\phi(K\phi_0) \, , 
\end{align*}
where we inserted the Schwinger-Dyson equation \eqref{eA.2} in the 
last step. Integrating this differential equation, we conclude that
\[
\no{F_\lambda} \ = \ \no{F} \mstern \ \, e^{i\lambda\phi(K\phi_0)} \, ,
\quad \lambda \in \RR \, .
\]
This relation is the integrated form of the Schwinger-Dyson 
equation in the perturbative setting.

\medskip 
We can proceed now to the time 
ordered exponentials of the functionals $\no{F}$ in 
the non-interacting theory with Lagrangean $L_o$,
$$
S_0(F) \, \doteq \  e_T^{\ i \, \no{\ F \ }}  \, .
$$
In case of nonlinear functionals,  
these time ordered exponentials are defined as 
formal power series with regard to Planck's constant. 
They can be interpreted as scattering matrices associated 
with the localized interaction induced by \mbox{$\no{F}$}. 
Noticing that 
\[
\delta L_0(\phi_0) = (L_0^{\phi_0} - L_0^{{ }}) = 
\phi(K \phi_0) + 1/2 \, \langle 
\phi_0, K \phi_0 \rangle \, , 
\]
where $K$ is the Klein Gordon operator, the Schwinger Dyson equation implies
\begin{equation} \label{eA.3}
S_0(F^{\phi_0}  + \delta L_0(\phi_0)) = S_0(F) \, \mstern \ \, 
S_0(\delta L_0(\phi_0)) = 
S_0(\delta L_0(\phi_0)) \, \mstern \ \, S_0(F) \, ;
\end{equation}
the second equality follows from the fact that the 
commutator function $\Delta$, appearing in the $\tstern$-product, is 
a bi-solution of the Klein-Gordon equation. 

\medskip
Next, we turn to the relative $S$-operators for interaction 
potentials $V$ of the type considered in this article. 
For given microcausal functional $F$ and any
test function $\chi$ which is equal to $1$ on $\supp F$,  
they are defined by
\[
S_{V(\chi)}(F) \doteq S_0(V(\chi))^{-1} \mstern \ \, S_0(V(\chi)+F) \, .
\]
Now let $\phi_0$ be any test function which also has support in 
the region where $\chi$ equals~$1$. Denoting by 
$L = L_0 + V$ the full Lagrangean for the given interaction
potential, we have
\begin{align*}
& S_{V(\chi)}(F) \, \mstern \ \, S_{V(\chi)}(\delta L(\phi_0)) \\
& = S_0(V(\chi))^{-1} \, \mstern \ \,   
S_0(V(\chi)+F) \, \mstern \ \, S_0(V(\chi))^{-1} \, \mstern \ \, 
S_0(V(\chi)+ \delta  L(\phi_0)) \, .
\end{align*}
Since $V(\chi)+ \delta L(\phi_0) = V(\chi)^{\phi_0} + \delta L_0(\phi_0)$, 
relation \eqref{eA.3} implies
\[
S_0(V(\chi)+ \delta L(\phi_0)) = S_0(V(\chi)^{\phi_0} + \delta L_0(\phi_0)) = 
S_0(V(\chi)) \ \mstern \ \, S_0(\delta L_0(\phi_0)) \, .
\]
It follows that
\[
S_{V(\chi)}(F) \mstern \ \, S_{V(\chi)}(\delta L(\phi_0))  = 
S_0(V(\chi))^{-1} \ \mstern \  S_0(V(\chi) + F) \ \mstern \ 
S(\delta L_0(\phi_0)) \, .
\]
Furthermore, the support properties 
of $\phi_0$ imply $\delta V(\chi)(\phi_0) = \delta V (\phi_0)$, hence 
\begin{align*}
& S_0(V(\chi)+F) \mstern \ \, S_0(\delta L_0(\phi_0)) \\ 
& = 
S_0(V(\chi)^{\phi_0} + F^{\phi_0} + \delta L_0(\phi_0)) 
= S_0(V(\chi)+ F^{\phi_0} + \delta L(\phi_0)) \, .
\end{align*}
Multiplying this equality from the left by 
$S_0(V(\chi)^{-1}$ finally gives
\begin{equation} \label{eA.4}
S_{V(\chi)}(F) \mstern \, S_{V(\chi)}(\delta L(\phi_0)) = 
S_{V(\chi)}(F^{\phi_0} \! + \! \delta L(\phi_0)) 
= S_{V(\chi)}(\delta L(\phi_0)) \mstern \, S_{V(\chi)}(F)  \, ,
\end{equation}
where the second equality follows from \eqref{eA.3} and 
the preceding relation. 
Since, the chosen test function $\chi$ is equal to $1$ on, both, the
supports of $F$ and $\phi_0$, we can omit this
spacetime cutoff of the potential. The equation resulting from
\eqref{eA.4} is the integrated Schwinger-Dyson equation in the presence 
of interaction.

\medskip 
Recalling that the $\tstern$-product corresponds to the
operator product, used in the main text, we have established
in the perturbative framework the dynamical relations, which were
taken as input in our C*-algebraic approach.
In a similar manner one can also justify the causal 
factorization relations.
Since these are widely discussed in the 
literature \cite{EG}, we refrain from reconsidering them here.

\medskip 
\vspace*{-1mm}
\section*{Acknowledgement}

\vspace*{-2mm}
DB gratefully acknowledges the hospitality extended to him by
Dorothea Bahns and the Mathematics Institute of the University of 
G\"ottingen.

\end{document}